\input harvmac
\input epsf

\Title{\vbox{\hbox{{\tt hep-th/0205266}}\hbox{KUCP0210}}}
{D5-brane in Anti-de Sitter Space and Penrose Limit}

\centerline{Shigenori Seki\footnote{$^\dagger$}{seki@phys.h.kyoto-u.ac.jp; 
The author will change the affiliation to 
Department of Physics, Faculty of Science, 
Kobe University, Kobe 657-8501, Japan on July 2002.} }
\bigskip
{\it \centerline{Department of Fundamental Sciences, FIHS}
\centerline{Kyoto University, Kyoto 606-8501, Japan}}

\vskip .3in

\centerline{{\bf abstract}}

We consider the Penrose limit of the solution of D5-brane given in the
Anti-de Sitter space and analyse the shape of the D5-brane in the pp-wave background. 
We find that the D5-brane leads to the branes and the throats connecting the branes. 
The branes spread on ${\bf R}^4$ with periodic values of the light-cone time $x^+$ and 
the throats lie along $x^+$. 
We also give some comments on holography.

\Date{May 2002}

\lref\BFHP{M. Blau, J. Figueroa-O'Farrill, C. Hull and G. Papadopoulos,
``A New Maximally Supersymmetric Background of IIB Superstring Theory'', 
JHEP 0201 (2002) 047, {\tt hep-th/0110242};
``Penrose Limits and Maximal Supersymmetry'', 
Class. Quant. Grav. 19 (2002) L87, {\tt hep-th/0201081}.}

\lref\BFP{M. Blau, J. Figueroa-O'Farrill and G. Papadopoulos,
``Penrose Limits, Supergravity and Brane Dynamics'',
{\tt hep-th/0202111}.}

\lref\BGMNN{D. Berenstein, E. Gava, J. Maldacena, K. S. Narain and H. Nastase,
``Open Strings on Plane Waves and their Yang-Mills Duals'',
{\tt hep-th/0203249}.}

\lref\BMN{D. Berenstein, J. Maldacena and H. Nastase, 
``Strings in Flat Space and PP Waves from ${\cal N}=4$ Super Yang Mills'',
JHEP 0204 (2002) 013, {\tt hep-th/0202021}.}

\lref\BN{D. Berenstein and H. Nastase,
``On Lightcone String Field Theory from Super Yang-Mills and Holography'',
{\tt hep-th/0205048}.}

\lref\CGS{C. G. Callan, A. Guijosa and K. G. Savvidy,
``Baryons and String Creation from the Fivebrane Worldvolume Action'',
Nucl. Phys. B547 (1999) 127, {\tt hep-th/9810092}.}

\lref\CGST{C. G. Callan, A. Guijosa, K. G. Savvidy and O. Tafjord
``Baryons and Flux Tubes in Confining Gauge Theories from Brane Actions'',
Nucl. Phys. B555 (1999) 183, {\tt hep-th/9902197}.}

\lref\CM{C. G. Callan and J. M. Maldacena,
``Brane Dynamics from the Born-Infeld Action'',
Nucl. Phys. B513 (1998) 198, {\tt hep-th/9708147}.}

\lref\DGR{S. R. Das, C. Gomez and S.-J. Rey,
``Penrose Limit, Spontaneous Symmetry Breaking and Holography in PP-wave Background'',
{\tt hep-th/0203164}.}

\lref\Imai{Y. Imamura,
``Supersymmetries and BPS Configurations on Anti-de Sitter Space'',
Nucl. Phys. B537 (1999) 184, {\tt hep-th/9807179}.}

\lref\KP{E. Kiritsis and B. Pioline,
``Strings in Homogeneous Gravitational Waves and Null Holography'',
{\tt hep-th/0204004}.}

\lref\LOR{R. G. Leigh, K. Okuyama and M. Rozali,
``PP-waves and Holography'',
{\tt hep-th/0204026}.}

\lref\Mal{J. M. Maldacena,
``The Large N Limit of Superconformal Field Theories and Supergravity'',
Adv. Theor. Math. Phys. 2 (1998) 231; Int. J. Theor. Phys. 38 (1999) 1113,
{\tt hep-th/9711200}.}

\lref\Met{R. R. Metsaev,
``Type IIB Green-Schwarz Superstring in Plane Wave Ramond-Ramond Background'',
Nucl. Phys. B625 (2002) 70, {\tt hep-th/0112044}.}

\lref\MT{R. R. Metsaev and A. A. Tseytlin,
``Exactly Solvable Model of Superstring in Plane Wave Ramond-Ramond Background'',
{\tt hep-th/0202109}.}

\lref\Pen{R. Penrose,
``Any space-time has a plane wave as a limit'',
Differential geometry and relativity, Reidel, Dordrecht (1976) pp. 271-275.}

\lref\Wit{E. Witten,
``Anti De Sitter Space and Holography'',
Adv. Theor. Math. Phys. 2 (1998) 253,
{\tt hep-th/9802150}.}

\lref\Witi{E. Witten,
``Baryons and Branes in Anti de Sitter Space'',
JHEP 9807 (1998) 006, {\tt hep-th/9805112}.}

\lref\Orb{
N. Itzhaki, I. R. Klebanov and S. Mukhi,
``PP Wave Limit and Enhanced Supersymmetry in Gauge Theories'',
JHEP 0203 (2002) 048, {\tt hep-th/0202153};
J. Gomis and H. Ooguri,
``Penrose Limit of N=1 Gauge Theories'',
{\tt hep-th/0202157};
M. Alishahiha and M. M. Sheikh-Jabbari,
``The PP-wave Limits of Orbifolded $AdS_5 \times S^5$'',
{\tt hep-th/0203018};
``Strings in PP-waves and Worldsheet Deconstruction'',
{\tt hep-th/0204174};
T. Takayanagi and S. Terashima,
``Strings on Orbifolded PP-waves'',
{\tt hep-th/0203093};
N. Kim, A. Pankiewicz, S.-J. Rey and S. Theisen,
``Superstring on PP-wave Orbifold from Large-N Quiver Gauge Theory'',
{\tt hep-th/0203080};
E. Floratos and A. Kehagias,
``Penrose Limits of Orbifolds and Orientifolds'',
{\tt hep-th/0203134};
K. Oh and R. Tatar,
``Orbifolds, Penrose Limits and Supersymmetry Enhancement'',
{\tt hep-th/0205067};
Y. Hikida and Y. Sugawara,
``Superstrings on PP-wave Backgrounds and Symmetric Orbifolds'',
{\tt hep-th/0205200}.
}

\lref\Bra{
A. Dabholkar and S. Parvizi,
``Dp Branes in PP-wave Background'',
{\tt hep-th/0203231};
A. Kumar, R. R. Nayak and Sanjay,
``D-Brane Solutions in PP-wave Background'',
{\tt hep-th/0204025};
D. Bak,
``Supersymmetric Branes in PP Wave Background'',
{\tt hep-th/0204033};
K. Skenderis and M. Taylor,
``Branes in AdS and PP-wave Spacetimes'',
{\tt hep-th/0204054};
M. Alishahiha and A. Kumar,
``D-Brane Solutions from New Isometries of PP-waves'',
{\tt hep-th/0205134}.
}

\lref\Oth{
J. G. Russo and A. A. Tseytlin,
``On Solvable Models of Type IIB Superstring in NS-NS and R-R Plane Wave Backgrounds'',
JHEP 0204 (2002) 021, {\tt hep-th/0202179};
M. Hatsuda, K. Kamimura and M. Sakaguchi,
``From Super-$AdS_5 \times S^5$ Algebra to Super-pp-wave Algebra'',
{\tt hep-th/0202190};
``Super-PP-wave Algebra from Super-$AdS \times S$ Algebras in Eleven-dimensions'',
{\tt hep-th/0204002};
M. Billo' and I. Pesando,
``Boundary States for GS Superstrings in an Hpp Wave Background'',
{\tt hep-th/0203028};
M. Cvetic, H. Lu and C. N. Pope
``Penrose Limits, PP-waves and Deformed M2-branes'',
{\tt hep-th/0203082};
``M-theory PP-waves, Penrose Limits and Supernumerary Supersymmetries'',
{\tt hep-th/0203229};
U. Gursoy, C. Nunez and M. Schvellinger,
``RG Flows from Spin(7), CY 4-fold and HK Manifolds to AdS, Penrose Limits and PP Waves'',
{\tt hep-th/0203124};
J. Michelson,
``(Twisted) Toroidal Compactification of PP-waves'',
{\tt hep-th/0203140};
S. R. Das, C. Gomez and S.-J. Rey,
``Penrose Limit, Spontaneous Symmetry Breaking and Holography in PP-wave Background'',
{\tt hep-th/0203164};
C.-S. Chu and P.-M. Ho,
``Noncommutative D-brane and Open String in PP-wave Background with B-field'',
{\tt hep-th/0203186};
J. P. Gauntlett and C. M. Hull,
``PP-waves in 11-dimensions with Extra Supersymmetry'',
{\tt hep-th/0203255};
P. Lee and J. Park
``Open Strings in PP-wave Background from Defect Conformal Field Theory'',
{\tt hep-th/0203257};
H. Lu and J. F. Vazquez-Poritz,
``Penrose Limits of Non-standard Brane Intersections'',
{\tt hep-th/0204001};
M. Spradlin and A. Volovich,
``Superstring Interactions in a PP-wave Background'',
{\tt hep-th/0204146};
Y. Imamura,
``Large Angular Momentum Closed Strings Colliding with D-branes'',
{\tt hep-th/0204200};
S. Frolov and A. A. Tseytlin,
``Semiclassical Quantization of Rotating Superstring in $AdS_5 \times S^5$'',
{\tt hep-th/0204226};
H. Takayanagi and T. Takayanagi,
``Open Strings in Exactly Solvable Model of Curved Spacetime and PP-wave Limit'',
{\tt hep-th/0204234};
I. Bakas and K. Sfetsos,
``PP-waves and Logarithmic Conformal Field Theories'',
{\tt hep-th/0205006};
C. Ahn,
``More on Penrose Limit of $AdS_4 \times Q^{1,1,1}$'',
{\tt hep-th/0205008};
``Comments on Penrose Limit of $AdS_4 \times M^{1,1,1}$'',
{\tt hep-th/0205109};
A. Parnachev and D. A. Sahakyan,
``Penrose Limit and String Quantization in $AdS_3 \times S^3$'',
{\tt hep-th/0205015};
H. Singh,
``M5-branes with 3/8 Supersymmetry in PP-wave Background'',
{\tt hep-th/0205020};
C. Kristjansen, J. Plefka, G. W. Semenoff and M. Staudacher,
``A New Double-Scaling Limit of N=4 Super Yang-Mills Theory and PP-wave Strings'',
{\tt hep-th/0205033};
N. R. Constable, D. Z. Freedman, M. Headrick, S. Minwalla, L. Motl, A. Postnikov and W. Skiba,
``PP-wave String Interactions from Perturbative Yang-Mills Theory'',
{\tt hep-th/0205089};
P. Bain, P. Meessen and M. Zamaklar,
``Supergravity Solutions for D-branes in Hpp-wave Backgrounds'',
{\tt hep-th/0205106};
S. D. Mathur, A. Saxena and Y. K. Srivastava,
``Scalar Propagator in the PP-wave Geometry Obtained from $AdS_5 \times S^5$'',
{\tt hep-th/0205136};
R. Gopakumar,
``String Interactions in PP-waves'',
{\tt hep-th/0205174};
K. Dasgupta, M. M. Sheikh-Jabbari and M. V. Raamsdonk,
``Matrix Perturbation Theory for M-theory on a PP-wave'',
{\tt hep-th/0205185};
L. A. P. Zayas and J. Sonnenschein,
``On Penrose Limits and Gauge Theories'',
{\tt hep-th/0202186};
G. Bonelli,
``Matrix Strings in PP-wave Backgrounds from Deformed Super Yang-Mills Theory'',
{\tt hep-th/0205213}.
}

\newsec{Introduction}

In the last several years, a lot of works have been done on the AdS/CFT 
correspondence \refs{\Mal,\Wit}, by which supersymmetric Yang-Mills theories 
are associated with string theories in Anti-de Sitter spaces.
When there are $N$ coincident D3-branes, we obtain an Anti-de Sitter space $AdS_5$
at the near-horizon limit.

Recently the string theory on a pp-wave background has been studied. 
The pp-wave background is known as one of maximally supersymmetric geometries.
The Type IIB string theory on this background can be exactly solved 
by the Green-Schwarz formalism 
in light-cone gauge \refs{\Met\MT\BFHP{--}\BFP}.
On the other hand, it has been pointed out that any space-time 
has a plane wave as a limit \Pen. This limit is called the Penrose limit.
Taking the Penrose limit for $AdS_5 \times S^5$, we obtain the pp-wave geometry.
It has been found that the spectra of the string theory on the pp-wave background
correspond to the operators of ${\cal N} = 4$ supersymmetric
Yang-Mills theory \refs{\BMN,\BGMNN}. A lot of related works have been done 
on orbifolds \Orb, D-branes \Bra, 
holography \refs{\DGR\KP\LOR{--}\BN}\ and so on \Oth.

In the AdS/CFT correspondence the baryon vertex in the world-volume theory 
on the $N$ coincident D3-branes is constructed by the D5-brane wrapped on $S^5$ \Witi.
The D5-brane surrounds the D3-branes and $N$ open strings connect 
these D-branes. This configuration is analysed in terms of 
the Born-Infeld action 
of D5-brane \refs{\CGS,\CGST}. The solution of the D5-brane has 
a spike sticking out toward the D3-branes and the tension of spike is equal to that of 
$N$ fundamental open strings.
The direction of the spike can be identified with the holographic direction. 
On the other hand, the holography in the pp-wave background 
has been studied in \refs{\DGR\KP\LOR{--}\BN}, but it has not been clear yet. 
Though the dual gauge theory of the string theory in pp-wave background has been found, 
we do not know where the dual theory lives, in other words, we do not have 
D-brane configurations as in the AdS/CFT correspondence.
In this paper we consider the Penrose limit of the solution of D5-brane given 
in the Anti-de Sitter space by the use of reasonable coordinate systems. 
We then analyse the shape of the D5-brane in the pp-wave background.

In Section 2 we introduce two types of coordinates 
of $AdS_5 \times S^5$ and 
confirm that the metrics written by such coordinates lead 
to the pp-wave metrics by the Penrose limit. 
In Section 3 the solution of D5-brane in $AdS_5 \times S^5$ is reviewed. 
We then calculate the Penrose limit of the solution and study 
the location of the D5-brane in the pp-wave background.
We find the branes connected by throats with each other.
Section 4 is devoted to conclusions and some comments on the holography.

\newsec{Anti-de Sitter space and Penrose limit}

The metric with $N$ coincident D3-branes has been known and 
it becomes $AdS_5 \times S^5$ at the near-horizon limit.
In the AdS/CFT correspondence we have often used the metric,
\eqnn\metads
\eqnn\rdef
$$\eqalignno{
ds^2 &= \left({u \over R}\right)^2 \left(-dt^2 + \sum_{i=1}^3dx_i^2\right) + 
\left({R \over u}\right)^2 du^2 + R^2d\Omega_5^2, &\metads\cr
&R^4=4\pi g_s N l_s^4 , &\rdef
}$$
for $AdS_5 \times S^5$. $g_s$ is a string coupling and $l_s$ is a string length.
But for the convenience of later analyses we rescale the coordinates and 
describe the metric as
\eqn\meti{
ds^2 = R^2 \left[l^2 \left( -dt^2 + \sum_{i=1}^3 dx_i^2 \right) + {1
\over l^2} dl^2\right] + R^2\left( d\theta^2 + \cos^2\theta d\psi^2 +
\sin^2\theta d{\tilde \Omega}_3^2\right).
}
In this coordinate system it is easy to find the descriptions of D-branes 
but it is hard to consider the Penrose limit. 
In order to make ready for taking the limit, we change the coordinates as
\eqnn\cort
\eqnn\corx
\eqnn\corl
$$\eqalignno{
t &= {\cosh \rho \sin \tau \over \cosh \rho \cos \tau - n_4 \sinh \rho},&\cort\cr
x_i &= {n_i \sinh \rho \over \cosh \rho \cos \tau - n_4 \sinh \rho},
\quad (i = 1, 2, 3) &\corx\cr
l &= \cosh \rho \cos \tau - n_4 \sinh \rho, &\corl\cr
&n_1^2 + n_2^2 + n_3^2 + n_4^2 = 1.
}$$
Substituting \cort, \corx\ and \corl\ into \meti, we obtain the metric 
\eqn\metii{
ds^2 = R^2(d\rho^2 - \cosh^2\rho d \tau^2 + \sinh^2 \rho d
\Omega_3^2) + R^2(d\theta^2 + \cos^2 \theta d\psi^2 + \sin^2\theta
d{\tilde \Omega}_3^2).
}
Now let us consider the Penrose limit. We reparametrize the coordinates as
\eqn\cortr{
\rho = {r \over R},\quad \tau = x^+ + {x^- \over R^2},\quad \theta =
{y \over R},\quad \psi = x^+ - {x^- \over R^2},
}
and take $R \to \infty$. The metric \metii\ then becomes 
\eqn\metiii{
ds^2 = -4dx^+dx^- - (r^2 + y^2)(dx^+)^2 + dr^2 + r^2 d\Omega_3^2 +
dy^2 + y^2d{\tilde \Omega}_3^2.
}
$(r, \Omega_3)$ and $(y, {\tilde \Omega}_3)$ describe ${\bf R}^4 \times {\bf R}^4$.
One of the two ${\bf R}^4$'s comes from $AdS_5$ and the other comes from $S^5$, but 
we can not distinguish between them at the Penrose limit. The radial coordinate of $AdS_5$,
which is associated with the holographic direction 
in the AdS/CFT correspondence,
leads to $r$. But $r$ and $y$ play same roles in the Penrose limit \LOR. 
So it is difficult to find the holographic direction in the limit. 
We will give some comments on this problem in the final section.

On the other hand, the metric \metiii\ can be identified with the pp-wave metric 
which includes Ramond-Ramond flux,
$$
F_{+1234} = F_{+5678}=1,
$$
where the suffices 1234 and 5678 denote the first and the second ${\bf R}^4$ respectively.

\newsec{Penrose limit of D5-brane}

Firstly we consider the D5-brane in $AdS_5 \times S^5$.
In order to find a baryon vertex \Witi, we set the D5-brane to wrap $S^5$.
The action of D5-brane is represented by Born-Infeld action and Chern-Simons action. 
The static solution of the action has been shown in \refs{\CGS,\CGST}. 
Let $(\xi_\alpha)$ be coordinates on the world-volume of D5-brane.
The action of D5-brane is described as
\eqnn\dvact
\eqnn\biact
\eqnn\csact
$$\eqalignno{
S &= S_{BI} + S_{CS} , &\dvact\cr
&S_{BI} = T_5\int d\xi^6 \sqrt{-\det (g + {\cal F})}, &\biact\cr
&S_{CS} = -T_5 \int A \wedge C^{(5)}, &\csact
}$$
where $T_5$ is the tension of D5-brane. 
$g$ is an induced metric, that is, 
$g^{\alpha\beta} = G^{ij}{\partial X_i \over \partial \xi_\alpha}
{\partial X_j \over \partial \xi_\beta}$, where $G^{ij}$ is a space-time metric.
${\cal F} (= \partial_\alpha A_\beta - \partial_\beta A_\alpha)$ is a gauge field strength 
on the D5-brane and $C^{(5)}$ is a five-form RR field strength.
We suppose that the D5-brane is embedded in $(t, \theta, \Omega_4)$ directions of 
the metric \meti.
We also set only $l$ and the gauge field $A_t$ to depend on $\theta$ \CGS.
$l$ is one of the coordinates denoting the location of D5-brane.
The action \dvact\ is written down as
$$
S = T_5 V_4 \int dt d\theta R^4 \sin^4\theta \left[-\sqrt{R^4(l^2 +
l'^2)- (\partial_\theta A_t)^2} + 4A_t \right] , 
$$
where $V_4$ is the volume of unit four-sphere and 
$'$ denotes the derivative by $\theta$. 
By the Legendre transformation and partial integral of the above action, we obtain 
the energy of D5-brane \refs{\CGS},
\eqnn\dvene
\eqnn\dth
$$\eqalignno{
E &= T_5 V_4 \int d\theta R^6\sqrt{l^2 + l'^2}\sqrt{f(\theta)^2 + \sin^8\theta}, &\dvene\cr
&f(\theta) \equiv \left[{3 \over 2}(\nu\pi - \theta) + {3 \over 2}
\sin\theta \cos\theta + \sin^3\theta \cos\theta\right] , &\dth
}$$
where $\nu$ is an integral constant and satisfies $0 \leq \nu \leq 1$.
From $\dvene$, we obtain the equation of motion for $l$,
\eqn\eneeom{
{d \over d\theta}\left[{l' \over \sqrt{l^2 + l'^2}}\sqrt{f(\theta)^2 +
\sin^8\theta}\right] = {l \over \sqrt{l^2 + l'^2}}\sqrt{f(\theta)^2 +
\sin^8\theta} .
}
On the other hand, the BPS condition of the D5-brane has been discovered by \Imai\ and 
 is denoted by
\eqn\eoml{
{l' \over l} = {\sin^5 \theta + f(\theta)\cos\theta \over
\sin^4\theta\cos\theta - f(\theta) \sin\theta}.
}
Note that \eoml\ is independent of $R$ because $f(\theta)$ does not include $R$.
We can confirm that the solutions of the BPS condition satisfy 
the equation of motion \eneeom.
Since \eoml\ is a first order differential equation, it is, in other words, the first integral of 
\eneeom. Though \eneeom\ seems to be too difficult to be solved, 
\eoml\ has been solved \refs{\CGS,\CGST} and the solution is
\eqn\soll{
l = {c \over \sin\theta}\left[{\theta - \pi\nu - \sin\theta \cos\theta
\over \pi(1-\nu)}\right]^{1 \over 3},
}
where $c$ is any constant.
We should note that \eneeom\ and \eoml\ have scale invariance for $l$, so 
we can introduce any scale parameter $c$.

Now let us consider the Penrose limits of the coordinates $l,t,x_i$ in \meti\ 
and make it clear the correspondences between the coordinates in \meti\ and in \metiii.
Substituting \cortr\ into \cort, \corx\ and \corl, we obtain
$$\eqalign{
t &= {\cosh {r \over R} \sin \left(x^+ + {x^- \over R^2}\right) \over
\cosh {r \over R} \cos \left(x^+ + {x^- \over R^2}\right) - n_4 \sinh
{r \over R}} , \cr
x_i &= {n_i \sinh {r \over R} \over \cosh {r \over R} \cos \left(x^+ +
{x^- \over R^2}\right) - n_4 \sinh {r \over R}} , \cr 
l &= \cosh {r \over R} \cos \left(x^+ + {x^- \over R^2}\right) - n_4 \sinh {r \over R} .
}$$
We can then calculate the limit $R \to \infty$ as 
\eqnn\ppt
\eqnn\ppx
\eqnn\ppl
$$\eqalignno{
t &\to \tan x^+ , &\ppt\cr 
x_i &\to 0, &\ppx\cr
l &\to \cos x^+. &\ppl
}$$
In $AdS_5 \times S^5$ the D3-branes, on which the dual gauge theory
exists, spread on the $(t, x_i)$ directions. \ppx\ implies that the space directions of 
the D3-branes shrink to zero size, 
while from \ppt\ we can read that the time direction 
of the D3-brane is included in the light-cone time $x^+$.
So the location of D3-brane in the pp-wave background is 
not clear. 

We consider the Penrose limit of the right hand side of \soll.
Substituting $\theta={y \over R}$ into \soll, we obtain
\eqn\solli{
l(y) = {c \over \sin{y \over R}}\left[{{y \over R} - \pi\nu -\sin{y
\over R}\cos{y \over R} \over \pi(1 - \nu)}\right]^{1 \over 3}.
}
When we take the limit $R \to \infty$, the two cases which are 
$\nu = 0$ and $\nu \neq 0$ give us different results.
Firstly we suppose that $\nu$ is equal to zero.
If $c$ is independent of $R$, the right hand side of \solli\ then converges on
$$
c\left({2 \over 3\pi}\right)^{1 \over 3}.
$$
We set that $c$ has a scale of $R^m$. If $m$ is negative, \solli\ goes to zero for that limit, 
while, if $m$ is positive, then \solli\ diverges. 
Note that the limit of \solli\ should take a value from $-1$ to $1$ on account of \ppl.
So the solutions of D5-brane with the scale $R^m$\ $(m>0)$ are not visible 
in the pp-wave background. For $m = 0$, that is, $c$ is independent of
$R$, we should take $|c| \leq \left({3\pi \over 2}\right)^{1 \over 3}$.
As a result the solutions of D5-brane in $x^+$-$y$ plane are
$$
x^+ = \pm {\rm constant} + 2k \pi,\quad k \in {\bf Z}.
$$

Next we suppose that $\nu \neq 0$. 
\solli\ converges on a non-trivial function by $R \to \infty$ 
only if $c ={a \over R}$, where $a$ is a constant independent of $R$.
We then obtain 
\eqn\limd{
l(y) \to {a \over y}\left({\nu \over \nu -1}\right)^{1 \over 3}.
}
Note that if $c$ has the scale of $R^m$ with $m < -1$, $l(y)$ shrinks to zero,
and that if $m > -1$, $l(y)$ goes away to infinity.
Let us analyse the case $c = {a \over R}$ in detail.
From \ppl\ and \limd, the solution of D5-brane at the Penrose limit
is described as
\eqn\dv{
\cos x^+ = {a \over y}\left({\nu \over \nu -1}\right)^{1 \over 3}.
}
\dv\ is still naive.
Since $y$ is a radial direction, $y$ should be positive. 
This condition can be satisfied by changing the sign of $a$.
Finally we obtain
\eqn\dvf{
y = \left({\nu \over 1 - \nu}\right)^{1 \over 3}\left|{a \over \cos x^+}\right| .
}
We should remember that $\nu$ is the integral constant satisfying 
$0 \leq \nu \leq 1$.
Since $y |\cos x^+| = |a|\left(\nu \over 1 - \nu\right)^{1 \over 3}$ 
from \dvf, 
at the limit $\nu \to 0$ we obtain $y = 0$ or $\cos x^+ = 0$.
These solutions can be rewritten by 
$$
y = 0, \quad x^+ = {\pi \over 2} + k \pi \quad (k \in {\bf Z}).
$$
On the other hand, let us take the limit $\nu \to 1$.
From the equation  ${1 \over y |\cos x^+|} 
= {1 \over a}\left(1 - \nu \over \nu\right)^{1 \over 3}$,
we obtain $y = \infty$ because $-1 \leq \cos x^+ \leq 1$.
\bigskip
\vbox{
\centerline{\epsfbox{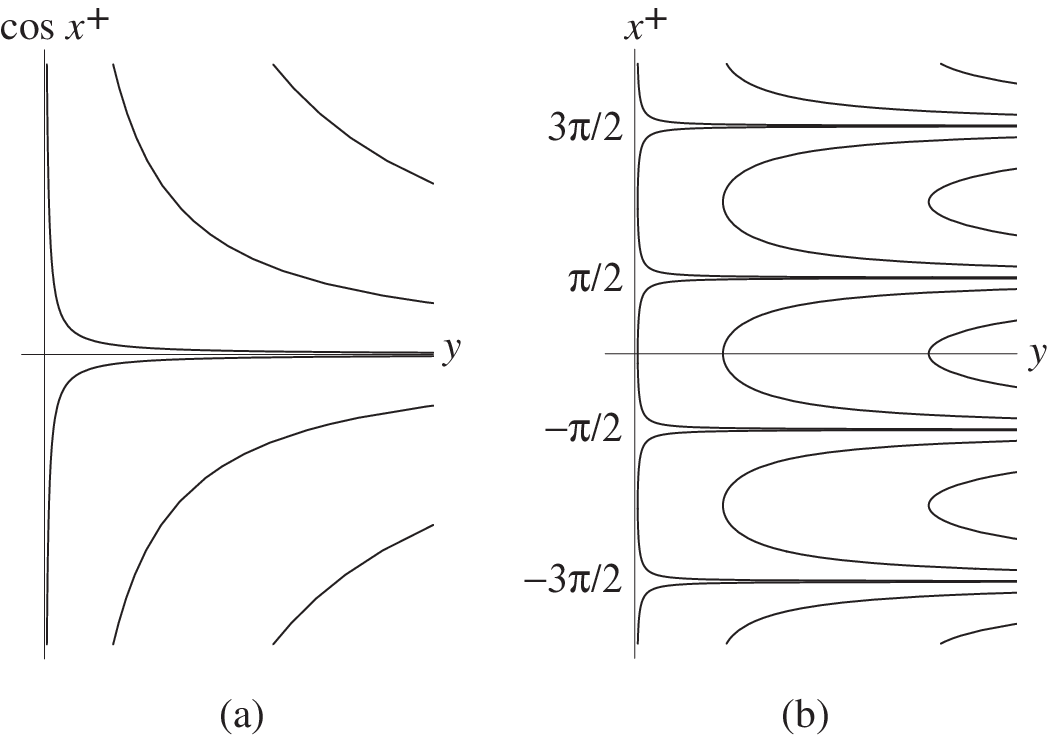}}
\vskip .5cm
\centerline{\fig\figi{The D5-brane in the Penrose limit}\ The Penrose limit of the D5-brane}
\centerline{\vbox{\vbox{(a) The D5-branes with three values of $\nu$ are
described on $y$-$\cos x^+$ plane and they}
\vbox{approach the axes when $\nu$ becomes smaller.}
\vbox{(b) The D5-branes are depicted on $y$-$x^+$ plane.}}}
}
\bigskip
The shape of the D5-brane is represented in \figi. When $\nu$ goes to zero,
the D5-brane approaches the axes in \figi\ (a).
The solution of D5-brane has periodicity for $x^+$ with period $\pi$.
From \figi\ (b) we can see that the D5-brane has throats at $y=0$.
When $\nu$ goes to zero, the throats are sharpened. 
Since the D5-brane was wrapped on the three-sphere ${\tilde \Omega}_3$ 
in $AdS_5 \times S^5$, 
it is also wrapped on the three-sphere ${\tilde \Omega}_3$ in the pp-wave background.
So at $x^+ = {\pi \over 2} + k \pi$ $(k \in {\bf Z})$ 
we obtain the branes spreading 
on ${\bf R}^4$ whose metric is given by 
$dy^2 + y^2 d{\tilde \Omega}_3^2$ in \metiii.
\bigskip
\vbox{\centerline{\epsfbox{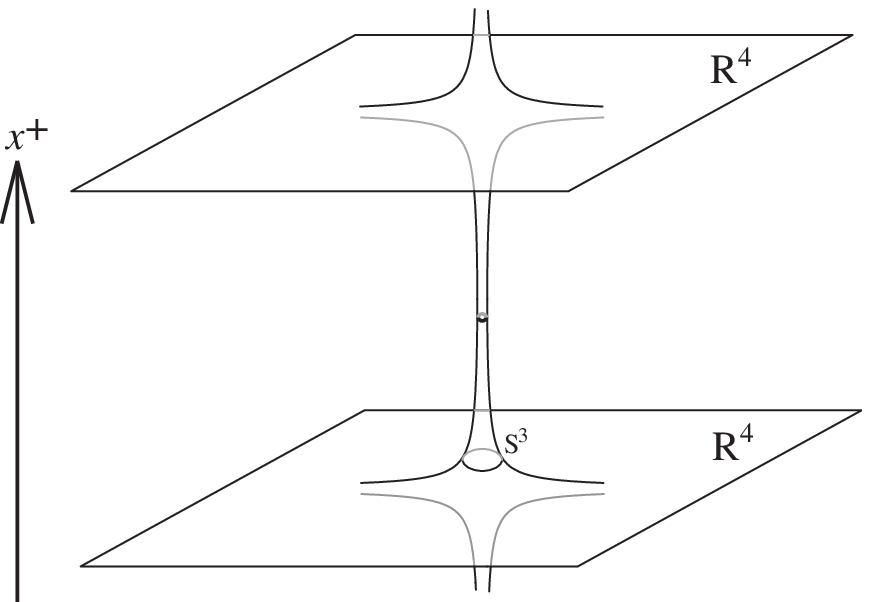}}
\vskip .5cm
\centerline{\fig\figii{D5-brane in the Penrose limit} Throats and branes as the Penrose limit of the D5-brane.}}
\bigskip
In \figii\ the branes lying on ${\bf R}^4$'s at $x^+ = \pm {\pi \over 2}$
are depicted. The throat connects these two branes. 

In the Anti-de Sitter space $\psi$ is periodic with period $2\pi$. 
Since $x^+$ is denoted by ${\tau + \psi \over 2}$, $x^+$ has $\pi$ periodicity.
That is the reason why we have found that the D-branes appear at the periodic values of 
$x^+$ with period $\pi$ in the pp-wave background.

\newsec{Conclusions and comments}

We have introduced mainly two types of the metrics for $AdS_5 \times S^5$. 
One is \meti, by which we can easily analyse the AdS/CFT correspondence. 
The solution of D5-brane in $AdS_5 \times S^5$ has been known \refs{\CGS,\CGST}.
The other is \metii, which is convenient for us to consider the Penrose limit.
But in this metric it is difficult to find the solution of D5-brane.
We have shown the relations between the two metrics and obtained the pp-wave metric
from the Penrose limit of \meti. 
Using this limit, we have analysed branes in the
pp-wave background. 
In $AdS_5 \times S^5$ the D5-brane has a spike sticking out along the
radial direction of $AdS_5$, while in the pp-wave metric we have found 
throats around $y=0$ and they lie along the light-cone time direction $x^+$.
At $x^+ = {\pi \over 2} + k \pi$ $(k \in {\bf Z})$ the brane spreads on the $y$ direction
and also wraps $S^3$ whose metric is $y^2d{\tilde \Omega}_3^2$ in \metiii.
So we can say that the brane lies on ${\bf R}^4$. 
This ${\bf R}^4$ is derived from a part of $S^5$ in $AdS_5 \times S^5$.
In the pp-wave metric there is the other ${\bf R}^4$ coming from a part of $AdS_5$.
Since there is no difference between these two ${\bf R}^4$'s, we would be 
able to find the brane wrapped on the latter ${\bf R}^4$. 

In the Penrose limit we have obtained the solution of D5-brane with
$\pi$ periodicity for $x^+$ direction. It is better that the
light-cone time $x^+$ is defined without a periodicity. 
In order to find non-periodic solutions of the branes, we should adopt
other choice of coordinate transformations \BN.

Though some proposals on the holography in the pp-wave background have
been done, we have not clearly known where the holographic direction
exists yet.
In \CM\ it is shown that the spikes sticking out from D-branes can be
identified with open strings. 
Let us remember \figii.
The D5-brane in the Anti-de Sitter space leads to the branes wrapping 
${\bf R}^4$'s with the fixed values of $x^+$ and the throats connecting 
the branes. The throats are sharpened as $\nu$ goes to zero.
We can regard the throats as open strings and they are ending on the 
D3-branes lying on ${\bf R}^4$. So we may be able to suggest that the 
holographic direction is $x^+$, which is transverse to the D3-branes.
But the correspondences between the gauge theory and the string theory in the pp-wave 
background may be essentially different from the AdS/CFT correspondence.
In order to make clear the holography in the pp-wave background, 
we need to consider the Born-Infeld action directly in the pp-wave metric and to 
find the solutions of D-branes.

\bigbreak\bigskip\bigskip\centerline{{\bf Acknowledgement}}\nobreak

I am grateful to K. Sugiyama for useful discussions.

\listrefs

\bye